\documentclass[aps,floatfix,nofootinbib,showpacs,twocolumn,superscriptaddress]{revtex4} 
 
\usepackage{amsmath} 
\usepackage{dcolumn} 
\usepackage{graphicx}

\begin{document} 
 
%\preprint{\parbox[t]{45mm}{\small ANL-PHY-12109-TH-2008}} 
 
\title{On unifying the description of meson and baryon properties}
 
\author{G.~Eichmann}
\affiliation{Physics Division, Argonne National Laboratory, 
             Argonne IL 60439-4843 U.S.A.} 
\affiliation{Institut f\"ur Physik, Karl-Franzens-Universit\"at Graz, A-8010 Graz, Austria}

\author{I.\,C.~Clo\"et} 
\affiliation{Physics Division, Argonne National Laboratory, 
             Argonne IL 60439-4843 U.S.A.} 
                          
\author{R.~Alkofer}
\affiliation{Institut f\"ur Physik, Karl-Franzens-Universit\"at Graz, A-8010 Graz, Austria}

\author{A.~Krassnigg}
\affiliation{Institut f\"ur Physik, Karl-Franzens-Universit\"at Graz, A-8010 Graz, Austria}

\author{C.\,D.~Roberts} 
\affiliation{Physics Division, Argonne National Laboratory, 
             Argonne IL 60439-4843 U.S.A.} 
             
\begin{abstract} 
%\rule{0ex}{3ex} 
A Poincar\'e covariant Faddeev equation is presented, which enables the simultaneous prediction of meson and baryon observables using the leading-order in a truncation of the Dyson-Schwinger equations that can systematically be improved.  
The solution describes a nucleon's dressed-quark core.  The evolution of the nucleon mass with current-quark mass is discussed.
A nucleon-photon current, which can produce nucleon form factors with realistic $Q^2$-evolution, is described.  
Axial-vector diquark correlations lead to a neutron Dirac form factor that is negative, with $r_1^{nu}>r_1^{nd}$.
The proton electric-magnetic form factor ratio falls with increasing $Q^2$.
\end{abstract} 
 
\maketitle 
 
\begin{figure}[b]
\centerline{%
\includegraphics[clip,width=0.45\textwidth]{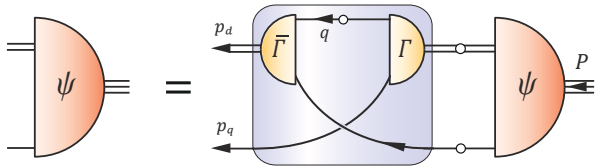}}
\caption{\label{faddeevfigure} %(Color online) 
Covariant Faddeev equation.  $\psi$: 
%in Eq.\,(\protect\ref{PsiNucleon}) 
Faddeev amplitude for nucleon of total momentum $P= p_q + p_d$.  It expresses the relative momentum correlation between the dressed-quark and -diquarks within the nucleon. Shaded region, equation's kernel: \emph{single line} --  dressed-quark propagator, $S(p)$, see Eq.\,(\protect\ref{rainbowdse}); $\Gamma$ -- diquark correlation amplitude, see Eq.\,(\protect\ref{bseD}); and \emph{double line} -- diquark propagator, see Fig.\,\protect\ref{qqprop}.
}
\end{figure}

%%%%%%%%%%%%%%%%%%%%%%%%%%%%%%%%%%%%%%%%%%%%%%%%%%%%%%%%%%%%%%%%%%%%%% 
In quantum field theory a nucleon appears as a pole in a six-point quark Green function.  The residue is proportional to the nucleon's Faddeev amplitude, which is obtained from a Poincar\'e covariant Faddeev equation that sums all possible exchanges and interactions that can take place between three dressed-quarks.  A tractable Faddeev equation for baryons was formulated in Ref.\,\cite{Cahill:1988dx}.  Depicted in Fig.\,\ref{faddeevfigure}, it is founded on the observation that an interaction which describes colour-singlet mesons also generates quark-quark (diquark) correlations in the colour-$\bar 3$ (antitriplet) channel \cite{Cahill:1987qr}.  While diquarks do not appear in the strong interaction spectrum; e.g., Refs.\,\cite{Bender:1996bb,Bender:2002as,Bhagwat:2004hn}, the attraction between quarks in this channel justifies a picture of baryons in which two quarks are always correlated as a colour-$\bar 3$ diquark pseudoparticle, and binding is effected by the iterated exchange of roles between the bystander and diquark-participant quarks.  

The study of baryons using the Poincar\'e-covariant dressed-quark Faddeev equation sits squarely within the ambit of the application of Dyson-Schwinger equations (DSEs) in QCD.  Notably, the DSEs admit a nonperturbative symmetry-preserving truncation scheme \cite{Bender:1996bb,Bender:2002as,Munczek:1994zz}, which has enabled the proof of numerous exact results for pseudoscalar mesons, and also provides for the formulation of reliable models that can be used to illustrate those results and make predictions for a wide range of meson observables with quantifiable errors; e.g., Ref.\,\cite{fn:review}.

Progress was recently made toward placing the computation of nucleon properties on the same level as that of mesons \cite{Eichmann:2007nn}.  In enabling the direct correlation of meson and nucleon properties via a single interaction kernel that preserves QCD's one-loop renormalisation group behaviour, it significantly complements existing phenomenological studies; e.g., Refs.\,\cite{Hecht:2002ej,Cloet:2008wg}.  Nevertheless, Ref.\,\cite{Eichmann:2007nn} can be refined.  Herein we report two material steps in that direction.

The first is to base the Faddeev equation on the amended one-parameter renormalisation-group-improved rainbow-ladder interaction introduced in Ref.\,\cite{Eichmann:2008ae}; i.e., with $\ell = k-q$,
%\begin{eqnarray}
%\nonumber \lefteqn{
%K^{tu}_{rs}(q,k;P) = - \,{\cal G}((k-q)^2) }\\
%
%&&  \times \, D_{\mu\nu}^{\rm free}(k-q)\,\left[\gamma_\mu \frac{\lambda^a}{2}\right]_{ts} \, \left[\gamma_\nu \frac{\lambda^a}{2}\right]_{ru} \!, \label{ladderK}
%\end{eqnarray}
\begin{equation}
K^{tu}_{rs}(q,k;P) = - \,{\cal G}(\ell^2)  D_{\mu\nu}^{\rm free}(\ell)\,\left[\gamma_\mu \frac{\lambda^a}{2}\right]_{ts} \, \left[\gamma_\nu \frac{\lambda^a}{2}\right]_{ru} \!, \label{ladderK}
\end{equation}
wherein: $r$,\ldots,\,$u$ represent colour and Dirac indices; $D_{\mu\nu}^{\rm free}$ is the free gauge boson propagator; and \cite{Eichmann:2008ae,Bloch:2002eq}
\begin{eqnarray}
\nonumber 
\frac{1}{Z_2^2}\frac{{\cal G}(s)}{s} &= & {\cal C}(\omega,\hat m) \, \frac{4\pi^2}{\omega^7} \,\frac{s}{\Lambda_t^4}\, {\rm e}^{-s/[\omega^2 \Lambda_t^2]}\\
&& + \frac{8\pi^2 \gamma_m}{\ln\left[ \tau + \left(1+s/\Lambda_{\rm QCD}^2\right)^2\right]} \, {\cal F}(s)\,, \label{Gkmodel}
\end{eqnarray}
with $Z_2$ the fermion wave function renormalisation constant, ${\cal F}(s)= [1-\exp(-s/\Lambda_t^2)]/s$, $\Lambda_t=1.0\,$GeV, $\tau={\rm e}^2-1$, $\gamma_m=12/25$, and $\Lambda_{\rm QCD} % = \Lambda^{(4)}_{\overline{MS}} 
= 0.234\,$GeV. In Eq.\,(\ref{Gkmodel}) 
\begin{equation}
\label{Cmodel}
\begin{array}{l}
\displaystyle {\cal C}(\omega,\hat m) ={\cal C}_0 +
\frac{{\cal C}_1(\omega-\bar\omega(\hat m))}{1 + {\cal C}_2 \,\hat x + {\cal C}_3^2\, \hat x^2},\;\hat x = \hat m / \hat m_0,\\[2.5ex]
{\cal C}_1(t) = 0.86 ( 1 - 0.15 t  + (1.50 t)^2 + (2.95t)^3)\,,
\end{array}
\end{equation}
with 
$\bar\omega(\hat m)=0.38 + 0.17/(1+\hat m/\hat m_0)$, where 
$\hat m$ is the renormalisation-group-invariant current-quark mass, $\hat m_0 = 0.12\,$GeV, and ${\cal C}_0=0.11$, ${\cal C}_2=0.885$, ${\cal C}_3=0.474$.  NB.\ By including dependence on $\hat m$, limited aspects of vertex corrections are represented in the kernel.

As explained in Ref.\,\cite{Eichmann:2008ae}, the interaction in Eq.\,(\ref{Gkmodel}) is characterised by a single constant; viz., ${\cal C}(\omega,\hat m)$.  It was determined subject to an understanding that corrections to the rainbow-ladder truncation which impact upon hadron phenomena, vanish with increasing current-quark mass; and in connection with light-quark systems, and those of the physical qualities of the pseudoscalar and vector mesons they constitute which are not tightly constrained by symmetries, the rainbow-ladder truncation of QCD's DSEs should produce results that, when measured in units of mass, are uniformly $\approx 35$\% too large.  The systematic implementation of corrections to this truncation then shifts calculated results so that reliable predictions and agreement with experiment can subsequently be obtained.  
(See, e.g., Refs.\,\cite{Hecht:2002ej,correct1,correct2,Watson:2004kd,Fischer:2007ze}.)
For each value of $\hat m$, computed \emph{observables} are approximately constant on the domain $\omega = \bar\omega(\hat m) \pm \delta\omega$, $\delta\omega=0.07$; i.e., $\delta\omega/\bar\omega(\hat m) \sim 0.2$.  

To define the Faddeev equation we %choose to 
compute the quark propagator in Fig.\,\ref{faddeevfigure} from the following gap equation
%The dressed-quark propagator in Fig.\,\ref{faddeevfigure} is computed from the rainbow gap equation; i.e., 
\begin{eqnarray}
\nonumber S(p)^{-1} & =&  Z_2 \,(i\gamma\cdot p + m^{\rm bm}) + \Sigma(p)\,,  \\
\Sigma_{tu}(p) & = & -\int^\Lambda_q\! K_{rs}^{tu}(q,p;P) S_{sr}(q) \,, \label{rainbowdse} 
\end{eqnarray} 
where $\int^\Lambda_q$ is a Poincar\'e invariant regularisation of the integral, with $\Lambda$ the regularisation mass-scale, and $m^{\rm bm}(\Lambda)$ is the current-quark bare mass.  $Z_{2}(\zeta^2,\Lambda^2)$ depends additionally on the renormalisation point, $\zeta$ and the gauge parameter.  
%
%The solution of Eq.\,(\ref{rainbowdse}) has the form 
%\begin{eqnarray} 
%\nonumber 
%S(p)^{-1} & = & i \gamma\cdot p \, A(p^2,\zeta^2) + B(p^2,\zeta^2) %\\ 
% 
%& =& \frac{1}{Z(p^2,\zeta^2)}\left[ i\gamma\cdot p + M(p^2,\zeta^2)\right] . 
%\label{sinvp} 
%\end{eqnarray}
%and is obtained subject to the condition $S(p)^{-1}|_{p^2=\zeta^2} = i\gamma\cdot p + m(\zeta)$,
%\begin{equation}
%\label{renormS} \left.S(p)^{-1}\right|_{p^2=\zeta^2} = i\gamma\cdot p + m(\zeta)\,,
%\end{equation}
%where $m(\zeta)$ is the renormalised mass: 
%\begin{equation}
%$Z_2(\zeta^2,\Lambda^2) \, m^{\rm bm}(\Lambda)$ $=$ $Z_4(\zeta^2,\Lambda^2) \, m(\zeta)$,
%\end{equation}
%with $Z_4$ the Lagrangian mass renormalisation constant.  
NB.\ We work in the isospin symmetric limit throughout.

The Faddeev equation requires diquark masses and correlation amplitudes.  They are obtained from the following Bethe-Salpeter equation \cite{Cahill:1987qr}:
\begin{equation}
\label{bseD}
\Gamma_{tu}^{qq}(k;P) = \int^\Lambda_l [\chi^{qq}(l;P)]_{sr}\, K_{us}^{tr}(l,k;P)\,,
\end{equation}
where: $k$ is the relative and $P$ the total constituent momentum; 
$\chi^{qq}(l;P):= S(l_+) \Gamma^{qq}(l;P) S(-l_-)^{\rm T}$ with ``T'' denoting matrix transpose, $l_\pm = l\pm P/2$; and $\Gamma^{qq}(k;P)$ is the correlation amplitude in a given colour-Dirac channel.  Equation~(\ref{bseD}) only has solutions in the $(\bar 3)_c$ channel so we can write $\Gamma^{qq} = H^a \Gamma_{J^P}^D$, where the colour is expressed through Gell-Mann matrices:
$\{H^1=i\lambda^7,H^2=-i\lambda^5,H^3=i\lambda^2\}$,
and $\Gamma_{J^P}^D$ is a Dirac-matrix-valued function whose explicit form depends on the $J^P$ (spin/parity) of the correlation. ($\bar\Gamma^{qq}(l;P) = C^\dagger \, \Gamma(-l;P)^{\rm T}\,C$, where $C=\gamma_2\gamma_4$ is the charge conjugation matrix.)

It has long been known that the lightest diquark correlations appear in the $J^P=0^+,1^+$
%scalar and axial-vector 
channels.  This can be seen \cite{Cahill:1987qr} to follow from the fact that $\Gamma^{D}_{J^P C}:= \Gamma_{J^P}^{qq}C^\dagger$ satisfies exactly the same Bethe-Salpeter equation as the $J^{-P}$ colour-singlet meson {\it but} for a halving of the coupling strength.  Hence only the $0^+$ and $1^+$ diquark correlations are retained in approximating the quark-quark scattering matrix to arrive at the Faddeev equation in Fig.\,\ref{faddeevfigure}.

\begin{figure}[t]
\vspace*{-8ex}

\centerline{\includegraphics[clip,scale=1.00]{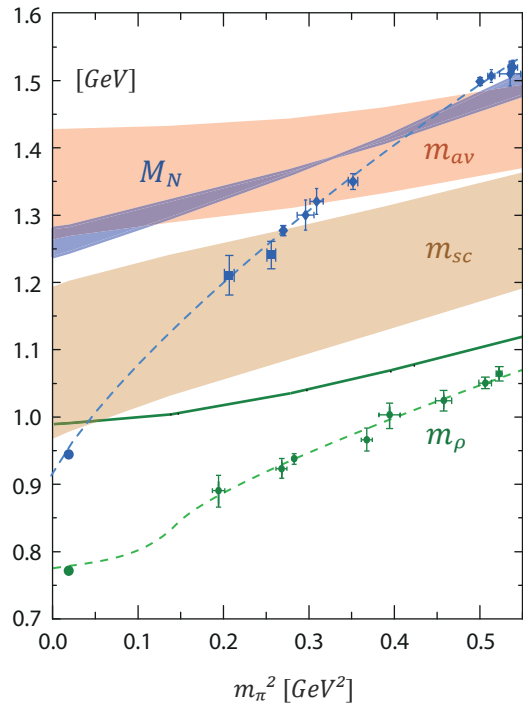}}

\vspace*{-4ex}

\caption[]{\label{fig:masses} 
%
%(Color online)
%
\emph{Thick bands}: Evolution with current-quark mass, $\hat m$, of the scalar and axial-vector diquark masses: $m_{sc}$ and $m_{av}$. Bands demarcate sensitivity to the variation $\omega = \bar\omega(\hat m) \pm \delta\omega$: lower (upper) edge corresponds to largest (smallest) $\omega$ value.  ($m_\pi$, calculated from rainbow-ladder meson Bethe-Salpeter equation: $\hat m = 6.1\,$MeV $\Rightarrow m_\pi = 0.138\,$GeV.)
\emph{Solid curve}: Evolution of $\rho$-meson mass \protect\cite{Eichmann:2008ae}.  %This observable quantity is insensitive to $\omega$.  
(NB.\ The interaction of Eqs.\,(\protect\ref{ladderK})--(\protect\ref{Cmodel}) was deliberately constructed to yield a calculated overestimate of $m_\rho(m_\pi^2)$.)
With $m_\rho$ we also depict results from simulations of lattice-regularised QCD \protect\cite{AliKhan:2001tx} along with an analysis and chiral extrapolation \protect\cite{Allton:2005fb}, \emph{short dashed curve}.  
\emph{Thin band}: Evolution with $\hat m$ of the nucleon mass obtained from our Faddeev equation: $\hat m =6.1\,$MeV, $M_N=1.26(2)\,$GeV.  For comparison we provide results from lattice-QCD \cite{Ali Khan:2003cu,Frigori:2007wa} and an analysis of such results \cite{Leinweber:2003dg}, \emph{dashed curve}.
} 
\end{figure}

In Fig.\,\ref{fig:masses} we present diquark masses computed with the interaction of Eqs.\,(\ref{ladderK})--(\ref{Cmodel}).  The bandwidths show their sensitivity to variations in $\omega$.  As diquarks do not appear in the strong interaction spectrum their masses are not observable.  
We note that the $\omega$-band on $m_{av}-m_{sc}$ is much narrower than that on the individual masses and that this difference falls with increasing current-quark mass.  
%
%As the $\Delta$-baryon may only involve axial-vector diquark correlations, $m_{av}-m_{sc}$ is a measure of the $\Delta$-$N$ mass splitting.  (At the physical pion mass $m_{av}-m_{sc} = 0.27(3)$ cf., experimentally, $M_\Delta - M_N=0.29$.)
%
Since the $\Delta$-baryon may only involve axial-vector diquark correlations, the $\Delta$-$N$ mass splitting is correlated with $m_{av}-m_{sc}$.  Hence we infer that $M_\Delta - M_N$ will depend weakly on $\omega$ and fall with increasing $m_\pi^2$.
Notwithstanding the correlation, near agreement between the experimental value of $M_\Delta - M_N=0.29\,$GeV and $m_{av}-m_{sc} = 0.27(3)\,$GeV at the physical pion mass is incidental \cite{Alkofer:2004yf}.  (NB.\ Hereafter parenthesised numbers indicate response to variation $\omega = \bar\omega(\hat m) \pm \delta\omega$.)
%
%(NB.\ At the physical pion mass $m_{av}-m_{sc} = 0.27(3)$ cf., experimentally, $M_\Delta - M_N=0.29$.)
%
%Our results indicate that $M_\Delta - M_N$ will depend weakly on $\omega$ and fall with increasing $m_\pi^2$.

To complete the Faddeev equation's kernel one must specify how the diquark correlations propagate.  This involves a definition of the diquark amplitudes off-shell and calculation of the diquark propagators.  For the former we observe that a diquark's correlation amplitude possesses one dominant piece; viz., that single Dirac-amplitude which would represent a point particle with the given quantum numbers in a local Lagrangian density, and a predetermined number of subdominant amplitudes.  In taking a diquark off-shell we require that for a correlation with large spacelike total momentum; i.e., $P^2 \gg M^2$, where $M$ is the relevant diquark mass, only the dominant amplitude remains pertinent.  
This is effected by noting that in any relevant tensor basis some of the subdominant amplitudes involve a factor of $\hat P$, where $\hat P^2=-1$ and $P$ is the total momentum of the correlation.  We multiply each $\hat P$ by a factor $h(x) = (1/i)(x/(x+2))^{1/2}$, where $x=P^2/M^2$.  Each subdominant amplitude is subsequently multiplied overall by a factor $g(x)=1/(x+2)$, while the Lorentz scalar functions associated with each matrix-valued tensor are fixed at the on-shell forms obtained in solving Eq.\,(\ref{bseD}).  (NB. On shell, $P^2 = -M^2$ and $h(-1)=1$, $g(-1)=1$.)  This is a simplification of the prescription detailed in Appendix~B of Ref.\,\cite{Eichmann:2007nn}.

\begin{figure}[b]
\begin{center}
\includegraphics[width=0.45\textwidth,clip]{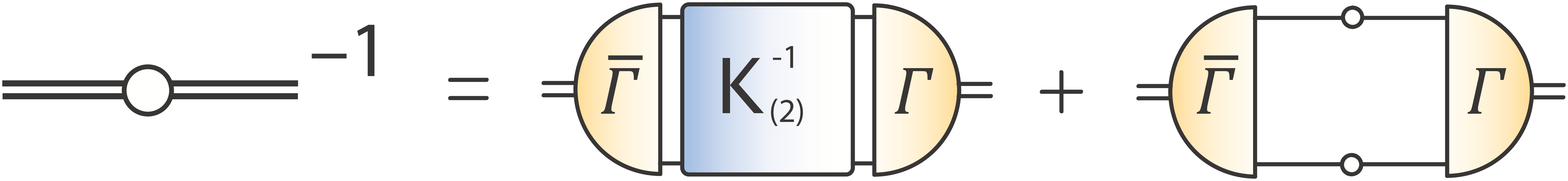}
\caption[]{\label{qqprop} Sum of diagrams that determine the diquark propagators in Fig.\,\protect\ref{faddeevfigure}.  $\Gamma$ represents the diquark correlation amplitudes; the single line, dressed-quark propagator; and ${\rm K}^{-1}$, inverse of the ladder-kernel in Eq.\,(\protect\ref{ladderK})--(\protect\ref{Cmodel}).}
\end{center}
\end{figure}

The propagators for the diquark correlations in Fig.\,\ref{faddeevfigure} can be written in the form ($x=P^2/M^2$)
\begin{eqnarray}
\label{propsc}
D_{sc}^{-1}(P^2) & = & M_{sc}^2 \, \left[ \Lambda_{sc} +\beta_{sc}\,{\cal F}(x) + {\cal Q}(x) \right], \\
\nonumber (D_{av}^{-1})_{\mu\nu}(P^2) & = & M_{av}^2\,\left[ \Lambda_{av}\,\delta_{\mu\nu}+\beta_{av}\,{\cal F}_{\mu\nu}(x) + {\cal Q}_{\mu\nu}(x)  \right].\\
\label{propav}
\end{eqnarray}
%They are calculated from the sum depicted in Fig.\,\ref{qqprop}, which has a form familiar from the bosonisation of non-local four-fermion interaction theories \cite{Cahill:1985mh}.  
%
On-shell the canonical Bethe-Salpeter normalisation condition ensures that in an internally consistent truncation: $\Lambda = -{\cal Q}_{(T)}(-1)$, $\beta = 1 - {\cal Q}^\prime_{(T)}(-1)$, ${\cal F}(-1)=0$, ${\cal F}^\prime_{(T)}(-1)=1$, where the subscript ``$(T)$'' is only relevant to the axial-vector propagator and identifies the transverse piece.
The second diagram is denoted by ``${\cal Q}$'' in Eqs.\,(\ref{propsc}), (\ref{propav}).  It is an easily evaluated, convergent one-loop integral whose integrand involves the dressed-quark propagator and the appropriate diquark correlation amplitude.  
The first diagram is associated with ${\cal F}$.  It represents a two-loop integral that must be computed on a large domain in the complex-$P^2$ plane.  In principle it can be evaluated using Monte-Carlo methods.  However, we circumvent that additional computer-time consuming step by employing the following parametrisations: 
${\cal F}(x) = [1+x/(x+2)^3]/4$; 
${\cal F}_{\mu\nu}(x) = \delta_{\mu\nu}\,[1- 1/(x+2)^2]/2$.
They are: based on an analytic analysis of the integrals; correct on-shell; and in qualitative accord with Monte-Carlo evaluations at real-$P^2$, with fair quantitative agreement at large spacelike momenta.

All elements in the Faddeev equation's kernel are now defined and computed.  (Its explicit form can be constructed following Sec.\,5 in Ref.\,\cite{Eichmann:2007nn} or App.\,A in Ref.\,\cite{Cloet:2008wg}.)  The equation is solved using standard methods.  Our result is depicted in Fig.\,\ref{fig:masses}.  Notably, despite the large $\omega$-dependence of the unobservable diquark masses, the nucleon mass is only weakly sensitive to this model parameter.  Our calculation directly correlates an efficacious one-parameter model of pseudoscalar and vector mesons \cite{fn:review} with the prediction of a nucleon observable.  
In particular, we have a systematically improvable continuum prediction for the evolution of the nucleon mass with a quantity that can methodically be connected with the current-quark mass in QCD.
%
%In particular we have the first continuum prediction of the evolution of the nucleon mass with a parameter that can be systematically connected to the current-quark mass in QCD.

We employed a kernel constructed carefully to ensure that corrections to the rainbow-ladder truncation systematically move results into line with experiment \cite{Eichmann:2008ae}.  The corrections can be divided into two classes: nonresonant diagrams and meson loops.  Both contribute materially at small $m_\pi^2$ but vanish with increasing current-quark mass.  
At the physical pion mass we predict $M_N=1.26(2)\,$GeV and hence that corrections to the rainbow-ladder truncation reduce the nucleon mass by $320\,$MeV.  A relevant comparison is provided by the analysis of lattice-QCD results in Ref.\,\cite{Young:2002cj}, which yields $M_N^\chi=:\alpha_N=1.27(2)\,$GeV.
Furthermore, it is important and internally consistent that at the physical pion mass $M_N/M_N^{\rm expt}=1.34$ and $m_\rho/m_\rho^{\rm expt} = 1.28$.  Likewise, it is significant that the difference between lattice results and our calculations diminishes with increasing current-quark mass.  

Our curve for the nucleon mass appears to lie below the lattice results at the largest current-quark masses.  This might indicate that our Faddeev kernel provides too much attraction.  That is conceivable given our replacement of the quark-quark scattering matrix by only the lightest correlations in the scalar and axial-vector diquark channels.  This is a truncation in addition to rainbow-ladder and its effect can be quantified.  In this connection, however, we note that the lattice length-scale (Sommer) parameter has decreased with the accessibility of lighter current-quarks \cite{McNeile:2007fu}.  Taking that into account for the $\rho$-meson trajectory would raise all of our calculated curves.  Therefore in our view Eqs.\,(\ref{ladderK})--(\protect\ref{Cmodel}) are satisfactory until the Sommer parameter stabilises.
% expt M_N/m_rho = 1.21
% calc. M_N/m_rho = 1.27
%At the largest current-quark mass in the figure we obtain a calculated ratio $M_N/m_\rho \approx 4/3$.

With the Faddeev amplitude in hand one can calculate nucleon form factors as a function of squared momentum transfer, $Q^2$.  The construction of a nucleon-photon vertex that fulfills the Ward-Takahashi identity for on-shell nucleons described by the Faddeev amplitude is detailed in Ref.\,\cite{Oettel:1999gc}.  In this approach one resolves the diquarks' substructure.  

Diquark correlations are not pointlike.  Hence, with increasing $Q^2$, diagrams in which the photon resolves a diquark's substructure must be suppressed with respect to contributions from diagrams that describe a photon interacting with a bystander or exchanged quark.  These latter are the only hard interactions with dressed-quarks allowed in a nucleon.  
This suppression was not expressed in the current of Ref.\,\cite{Eichmann:2007nn} and that leads to spurious results; e.g., the proton electric form factor does not approach zero with increasing $Q^2$.  
Here we introduce a second, major improvement of Ref.\,\cite{Eichmann:2007nn}; viz., we refine the nucleons's electromagnetic current and alter as described below the expressions in App.\,D of Ref.\,\cite{Eichmann:2007nn}.  

A key modification is to follow Ref.\,\cite{Eichmann:2008ae} and write the dressed-quark-photon vertex as ($y=Q^2/m_\rho^2$)
%\begin{eqnarray}
%\nonumber\lefteqn{\Gamma^\mu(k;Q) = \Gamma^\mu_\text{BC}(k;Q)}\\
%
%&& - \frac{f_\rho}{m_\rho} \,\frac{y}{y+1} \,T_{\mu\nu}^Q \,\Gamma^\nu_{vc}(k;Q)\,
%                e^{-(\rho_1 + \rho_2 \,y^2)\,(1+y)},
%\label{qgvertex}
%\end{eqnarray}
\begin{equation}
\Gamma^\mu(k;Q) = \Gamma^\mu_\text{BC}(k;Q) -\frac{f_\rho}{m_\rho} \,\frac{y}{y+1} \,T_{\mu\nu}^Q \,\Gamma^\nu_{vc}(k;Q)\, e^{-g(y)}.
\label{qgvertex}
\end{equation}
Here the first term is the regular Ball-Chiu \emph{Ansatz} \cite{Ball:1980ay,Roberts:1994hh}.  The second introduces a resonant $\rho$-meson contribution, which is explicitly excluded from the impulse current when one employs a Ball-Chiu vertex calculated from the solution of the rainbow gap equation \cite{Eichmann:2007nn}.  In this piece: $f_\rho$ is the calculated on-shell $\rho$-meson leptonic decay constant \cite{Ivanov:1998ms}; $T_{\mu\nu}^Q$ is a projector transverse to the four-vector $Q$; and $\Gamma^\nu_{vc}(k;Q)$ is the calculated and canonically normalised rainbow-ladder $\rho$-meson Bethe-Salpeter amplitude, whose extension off-shell follows the pattern outlined above for diquark correlations.  The factor involving $g(y)=(\rho_1 + \rho_2 \,y^2)\,(1+y)$, with $\rho_{1,2}$ parameters, accounts for any additional effect owing to an off-shell $\rho$-meson.  Emulating Ref.\,\cite{Eichmann:2008ae}, $\rho_1$ was determined by requiring that Eq.\,(\ref{qgvertex}) reproduce the DSE $\pi$-meson charge radius curve in Fig.\,6 of Ref.\,\cite{Maris:2005tt}: $\rho_1^2 = 0.001+m_\pi^2/(3.72\,{\rm GeV})^2$.  $\rho_2$ is relevant to the $Q^2$-evolution of form factors.

To ensure that a diquark's contribution to the nucleons' form factors is suppressed with increasing $Q^2$ it is necessary to modify the so-called seagull terms in the current.  If the seagull vertices employed in Ref.\,\cite{Eichmann:2007nn} are denoted generically by $M^{WT}_\mu\!$, then a minimal correction is provided by
\begin{equation}
\label{seagullmod}
%M_\mu = [1 - m(y)] M_\mu^{WT} 
M_\mu = M_\mu^{WT} -  m(y)T_{\mu\nu}^Q M_\nu^{WT},
\end{equation}
where $m(y)$ vanishes at $y=0$ and as $y\to\infty$ but is otherwise constrained little. Since a $\rho$-meson contribution to the quark-photon vertex will be communicated to the seagull terms in any internally consistent calculation of these five-point Schwinger functions, we employ
\begin{equation}
m(y) = \frac{1}{\surd 2} \frac{f_\rho}{m_\rho} \, \frac{y^2}{1+y} \, {\rm e}^{-\rho_3 (1+y)},
\label{seagull}
\end{equation}
with $\rho_3$ a parameter that governs the degree of suppression with increasing $Q^2$. $m(y)$ in Eq.\,(\ref{seagullmod}) has no effect on the nucleon's static electromagnetic properties.  

\begin{figure}[t]
\centerline{%
\includegraphics[clip,width=0.45\textwidth]{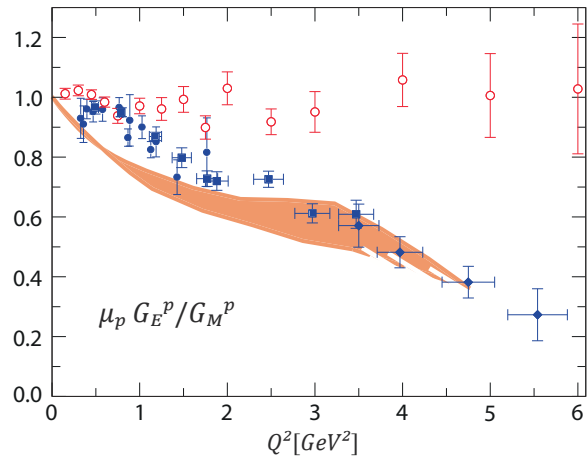}}

\caption{\label{GEGMproton} 
%
%(Color online)
%
\emph{Shaded band} -- Calculated result for $\mu_p G_E^p(Q^2)/G_M^p(Q^2)$.  Bandwidth delimits the sensitivity to $\omega$ on the domain $\omega = \bar\omega(\hat m) \pm \delta\omega$: larger (smaller) $\omega$, lower (upper) border.  Data: \emph{open circles}, Ref.\,\protect\cite{walker}; \emph{filled squares},  Ref.\,\protect\cite{jones}; \emph{filled circles}, Ref.\,\protect\cite{roygayou}; and \emph{filled diamonds}, Ref.\,\protect\cite{gayou}.}
\end{figure}

\begin{table}[b]
\caption{\label{static} Static properties calculated with $\hat m = 6.1\,$MeV: magnetic moments in magnetons defined by calculated nucleon mass, $M_N$.
%; radii length scale is fm.  
%Parenthesised numbers indicate response to variation $\omega = \bar\omega(\hat m) \pm \delta\omega$, Eq.\,(\protect\ref{Cmodel}).  
%
\emph{Row 2}: Results determined by Ref.\,\protect\cite{Cloet:2008wg}.
\emph{Row 3}: Experimental \protect\cite{Yao:2006px} or inferred \protect\cite{Hammer:2003ai} values.}
\begin{ruledtabular} 
\begin{tabular*} {\hsize} 
{l@{\extracolsep{0ptplus1fil}} |
l@{\extracolsep{0ptplus1fil}}
l@{\extracolsep{0ptplus1fil}}
l@{\extracolsep{0ptplus1fil}}
l@{\extracolsep{0ptplus1fil}}
l@{\extracolsep{0ptplus1fil}}
l@{\extracolsep{0ptplus1fil}}} 
%
%\\\hline
& $\mu_n$ & $\mu_p$ & $(M_n r_E^n)^2$ & $M_p r_E^p$ & $M_n r_M^n$ & $M_p r_M^p$  \\\hline  
Calc.\ & -1.58(3) & 2.56(5) & ~0.00(23) & 3.76(5) & 3.43(5) & 3.47(5)\\
Ref.\protect\cite{Cloet:2008wg} & -1.51 & 2.55 & -0.60 & 3.41 & 3.05 & 2.99\\
Expt.\ & -1.91 & 2.79 & -2.63 & 4.16 & 4.19 & 4.07 \\
\end{tabular*} 
\end{ruledtabular} 
\end{table} 
%    & $\mu_n$ & $\mu_p$ & $(r_E^n)^2$ & $r_E^p$ & $r_M^n$ & $M r_M^p$ \\\hline
%Calc.\ & -1.58(3) & 2.56(5) & ~0.00(1) & 0.79(1) & 0.72(1) & 0.73(1)\\
%
%Ref.\protect\cite{Cloet:2008wg} & -1.51 & 2.55 & -0.02 & 0.57 & 0.51 & 0.50\\
%
%Expt.\ & -1.91 & 2.79 & -0.12 & 0.88 & 0.88 & 0.86 \\
%

Calculation of the nucleons' electromagnetic form factors is now a straightforward numerical exercise.  The results correlate a one-parameter model of pseudoscalar and vector meson properties with the prediction of a range of nucleon observables.  A raft of results will be reported elsewhere.  Herein we describe only the static electromagnetic properties, listed in Table~\ref{static}, and the proton form factor ratio, depicted in Fig.\,\ref{GEGMproton}.  

Regarding static properties, the $\rho$-meson piece of Eq.\,(\ref{qgvertex}) provides $\sim 50$\% of $r_E^2$, as was also the case for the pion \cite{Eichmann:2008ae}.  Our kernel deliberately omits meson cloud contributions, whose effect was discussed most recently in Ref.\,\cite{Cloet:2008wg}.  They contribute uniformly to improving agreement with experiment in Table~\ref{static}.  In accord with the model of Ref.\,\cite{Cloet:2008wg}, we predict quark-core values: $\mu_p^u=2.36(5)$, $\mu_p^d=0.20(1)$, where $\mu_p^f$ means the contribution from a quark of flavour $f$ to the proton's magnetic moment.  Charge symmetry entails $\mu_n^u=-0.40(1)$, $\mu_n^d=-1.18(3)$.  NB.\ At each current-quark mass considered, Fig.\,\ref{fig:masses}, no calculated magnetic moment or charge radius varies by more than 4\% on the domain $\omega = \bar\omega(\hat m) \pm \delta\omega$.

Of interest is the behaviour of the neutron's Dirac form factor.  One can write $F_1^n(Q^2)= (2/3)[ f_1^{nu}(Q^2) - f_1^{nd}(Q^2)]$. %, where $f_1^{nf}(Q^2)$ is the contribution to $F_q^n$ from a quark of flavour $f$.  
Our framework predicts $F_1^n(Q^2)<0$ with an evolution at small-$Q^2$ that can be characterised by two Dirac radii; namely, $r_1^{nu} = 0.85(2)\,$fm, $r_1^{nd}=0.76(1)\,$fm. That our calculated $u$-quark contribution to $F_1^n$ evolves more rapidly than that of the $d$-quark is consistent with contemporary parametrisations of experimental data; e.g., Ref.\,\cite{Kelly:2004hm}.  It owes to the presence of axial-vector diquark correlations in our Faddeev amplitude.

The neutron's Faddeev amplitude involves the quark-diquark flavour structures $d[ud]_{0^+}$ and $\{\surd 2 u (dd)_{1^+} - d (ud)_{1^+}\}$, with the axial-vector correlation being more massive than the scalar.  The Faddeev equation ensures that the two $d$-quarks are each equally likely to be found in a diquark correlation.  Hence, the $ud$ correlations do not favour localisation of the $u$-quark at the neutron's centre-of-mass any more than they do the $d$-quark.   However, a localisation of the $d$-quarks does follow from the significant probability that the neutron's two $d$-quarks are embedded in a massive $1^+$ diquark correlation.  
(NB.\ We have assumed isospin symmetry and hence properties of $F_1^p$ can readily be inferred from this discussion.)

The proton form factor ratio is depicted in Fig.\,\ref{GEGMproton}.  Our truncation omits pseudoscalar meson cloud contributions.  We therefore have $r_E^p>r_M^p$ and disagree with experiment for $Q^2\lesssim 2\,$GeV$^2$.  (This also explains our inadequate description of $r_E^n$.)  Since pseudoscalar mesons are not pointlike, such contributions rapidly become unimportant beyond this scale and hence the quark core described by our Faddeev equation should be quantitatively reliable.  (See Sec.~4.2.3 of Ref.\,\cite{Alkofer:2004yf}.)  We report $\mu_p G_E^p(Q^2)/G_M^p(Q^2)$ because its $Q^2 >2\,$GeV$^2$ behaviour is sensitive to the nucleon-photon current and therefore Eqs.\,(\ref{qgvertex})--(\ref{seagull}) become important \emph{on that domain}.  We find $(\rho_{2},\rho_3)=(0.001,0.075)$ optimises agreement with the polarisation transfer data \cite{jones,roygayou,gayou}.  Variations of $\pm 20$\% have a noticeable impact but no physically acceptable choice for $\rho_{2,3}$ can reproduce the Rosenbluth data \cite{walker}.  This is the only parameter fitting herein.  
%$\mu_p G_E^p(Q^2)/G_M^p(Q^2)=\,$constant.
NB.\ Calculating the dressed-quark-photon vertex would obviate the need for the \textit{Ans\"atze} and parameters $\rho_{1,2,3}$ in Eqs.\,(\ref{qgvertex})--(\ref{seagull}) and is a natural next step toward true predictions of the large-$Q^2$ behaviour of nucleon form factors.

Our description of diquark propagation is inadequate for $Q^2\gtrsim 4\,$GeV$^2$ whereupon its failure to implement diquark confinement leads to pinch singularities within the form factor integration domain.  This defect is absent in the phenomenological models of Refs.\,\cite{Hecht:2002ej,Cloet:2008wg,Alkofer:2004yf} and must be circumvented before computation at larger $Q^2$ can proceed within our \emph{ab-initio} framework.
NB.\ Even with a realisation of confinement the large-$Q^2$ form factor integration domain might contain moving singularities \cite{Bhagwat:2002tx}.  An algorithm for handling that case is also lacking.
%
%It is striking that a parameter-free Faddeev equation calculation, constrained solely by meson physics, should yield a result for the ratio that is in agreement with the polarisation transfer data \cite{jones,gayou}.

We constructed a parameter-free Faddeev equation whose solution describes a nucleon's dressed-quark core.  Its kernel is built from a renormalisation-group-improved, current-quark-mass-dependent rainbow-ladder interaction that provides a sound description of pseudoscalar and vector mesons and, in particular, a veracious description of the pion as both a Goldstone mode and a bound state of dressed-quarks \cite{fn:review}.  This enables the simultaneous calculation of baryon and meson properties using a well-defined truncation of the DSEs that can systematically be improved.  We also presented a nucleon-photon current that automatically preserves the Ward-Takahashi identity for on-shell nucleons described by the Faddeev amplitude we obtained.  It can produce nucleon form factors with realistic $Q^2$-evolution.

%------------------------------------------------------------------------ 
%\begin{acknowledgments} 
%
%We acknowledge: conversations with B.~El-Bennich, T.~Kl\"ahn, D.~Nicmorus and R.\,D.~Young;
%
%Department of Energy, Office of Nuclear Physics, contract no.\ DE-AC02-06CH11357; 
%
%Austrian Science Fund (\emph{FWF}) grant no.\ W1203 
%(Doctoral Program ``Hadrons in vacuum, nuclei and stars'') 
%\& project nos.\ P20496-N16, P20592-N16;
%
%and use of ANL's Computing Resource Center's facilities.
%
We acknowledge helpful conversations with B.~El-Bennich, T.~Kl\"ahn,
D.~Nicmorus and R.\,D.~Young, and thank P.\,C.~Tandy for a critical
reading of the manuscript.  We acknowledge support from
Department of Energy, Office of Nuclear Physics, contract no.\
DE-AC02-06CH11357,
Austrian Science Fund (\emph{FWF}) grant no.\ W1203
%(Doctoral Program ``Hadrons in vacuum, nuclei and stars'')
\& project nos.\ P20496-N16, P20592-N16;
and use of ANL's Computing Resource Center's facilities.
%
%\end{acknowledgments} 
%------------------------------------------------------------------------ 

\end{document}